\documentclass[a4paper,11pt]{article}
\usepackage{pos}

\title{Unraveling QCD dynamics via charm-tagged jets with ALICE at the LHC}

\author[a]{Emma Yeats}

\affiliation[a]{University of California Berkeley,\\
  Berkeley, California, United States}

\emailAdd{eyeats@berkeley.edu}

\abstract{The study of jet substructure provides tests of Quantum Chromodynamics (QCD) and offers differential ways to study hadronization mechanisms. QCD predicts that jet radiation patterns depend on the mass and color charge of the initiating parton. Parton showers are sensitive to the different Casimir factors of quarks and gluons, and to the parton mass through the dead-cone effect. 
In this talk, we present three sets of $\textrm{D}^0$-meson-tagged jet measurements. First, we present the first measurement of the Energy-Energy Correlator (EEC) for $\textrm{D}^0$-meson-tagged jets. This observable reveals the detailed structure of QCD radiation from heavy quarks, allowing for a distinction between perturbative and non-perturbative effects in charm-quark showers. We then discuss measurements of the angular differences between three different axis definitions inside the jets. The three definitions offer different sensitivity to wide-angle soft radiation and systematically probe the radiation pattern from charm quarks. The final measurement regards jet angularities, which characterize the weighted angular distribution of the jet constituents. These powerful observables probe the parton shower, with sensitivity to parton-mass and Casimir-factor related effects that can be tuned by adjusting the importance of wide-angle radiation.}

\FullConference{43rd International Conference on High Energy Physics (ICHEP 2026)\\
30 July  to 5 August , 2026\\
Natal, Brazil\\}

\begin{document}
\maketitle

\section{Introduction}
Heavy quarks are predominantly produced in the initial hard scattering in proton--proton (pp) collisions, and can be traced to the final-state hadrons observed in detectors. Jets tagged with heavy-flavor quarks are excellent probes for testing perturbative QCD calculations, and have a distinct radiation pattern in pp collisions due to the presence of two QCD phenomena.
First, a difference in Casimir color factors between quarks and gluons leads to gluon-initiated showers exhibiting a broader and softer fragmentation profile than quark-initiated showers, which have a more concentrated jet core and generally fewer splittings.
The second is the dead-cone effect, which predicts that gluon emissions are suppressed in a cone around the direction of a radiating heavy quark \cite{theorydeadcone}. The size of this cone is proportional to the radiating quark's mass and inversely proportional to its energy, $\theta_{\textrm{dead-cone}} = m_{\textrm{radiator}}/E_{\textrm{radiator}}$, and thus should be most prominent in the low charged-jet transverse momentum ($p_{\textrm{T, ch jet}}$) regime commonly studied by the ALICE experiment. The dead-cone effect was first observed by ALICE in 2022 \cite{deadcone}, in 5 < $p_{\textrm{T, radiator}}$ < 10 GeV/$c$ charm jets, and was subsequently observed in beauty jets by both the LHCb and CMS experiments \cite{lhcbdeadcone, cmsdeadcone}.

We can study the interplay between these two shower phenomena with substructure measurements in jets tagged with a heavy quark. Charm jets are tagged by the presence of a D$^0$ meson reconstructed from the decay of D$^0$ $\rightarrow$ K$^- \pi^+$ (and charge conjugate), which has a branching ratio of ($3.947 \pm 0.030$)\% \cite{PDG}. Pairs of daughter tracks passing topological selections are replaced by the corresponding D$^0$ candidate, whose four-momentum is taken as the sum of the daughters' four-momenta. Charged jets are reconstructed with the anti-$k_{\textrm{T}}$ algorithm \cite{anti-kT} with radius $R=0.4$ for all measurements discussed. Unless otherwise stated, the $E-$recombination scheme is used to define the jet axis. We report three recent D$^0$-tagged jet substructure measurements from the ALICE experiment, which are IRC safe to allow for comparison with precise theoretical predictions.

\section{$\textrm{D}^0$-tagged energy-energy correlator at $\sqrt{s}=13$ TeV}\label{sec:eec}

The EEC is the energy-weighted two-particle correlation inside jets \cite{eec} and is given by:
\begin{equation}
\Sigma_{\rm EEC}(R_{\rm L}) = \frac{1}{N_{\rm jet}\Delta}\sum_{N_{\rm jet}}\int\sum_{i,j} \left( \frac{p_{{\rm T}, i} p_{{\rm T}, j}}{p_{\rm T, jet}^2} \right) \delta(R_{\rm L}' - R_{{\rm L}, ij}){\rm d}R_{\rm L}',
\end{equation}
where the summation runs over the total number of jets, $N_{\rm jet}$, and all final-state particle pairs ($i,~j$) inside each jet. The angular interval size is denoted by $\Delta$. The term in the parentheses is an energy-weight, which utilizes the transverse momentum of the pairs over $p_{\rm T, jet}^2$ to suppress the contribution of soft particles. To construct the EEC, the opening angle between each pair is taken as $R_{{\rm L}} = \sqrt{\Delta \varphi_{ij}^2 + \Delta \eta_{ij}^2}$. Pairs that are further apart are correlated to earlier splittings in the jet shower history, and pairs that are closer together correlate to splittings that occur at later times. Thus, the EEC is theorized to separate perturbative (large-$R_{\textrm{L}}$) from non-perturbative physics (small-$R_{\textrm{L}}$). The transition region, or ``peak", of the EEC corresponds to the onset of hadronization and is sensitive to the mass of the quark initiating the jet \cite{kyle_lee_EEC}.

\begin{figure}
\centerline{\includegraphics[width=0.5\linewidth]{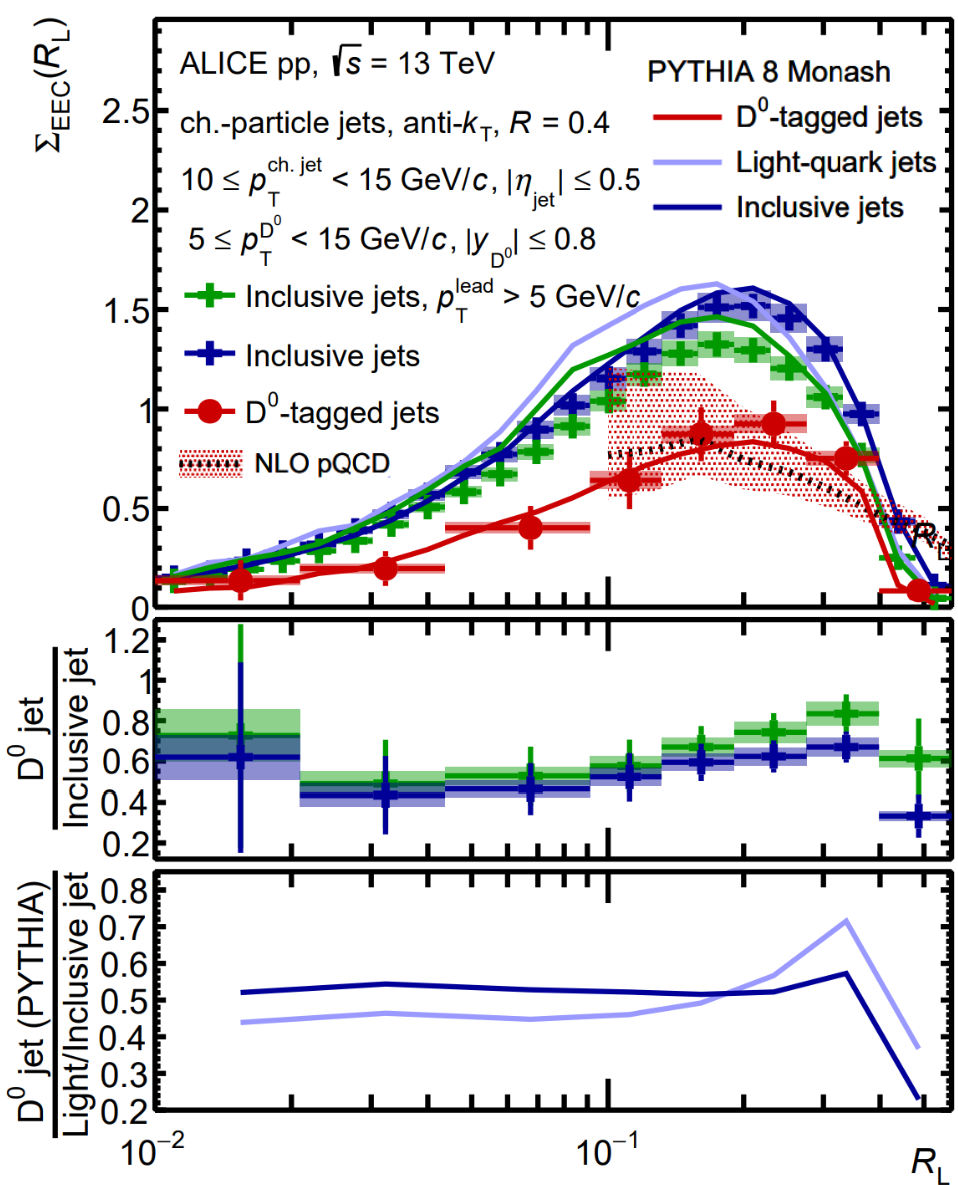}}
\caption[]{The $\rm D^0$-tagged jet EEC distribution (red) compared to inclusive jets, with (green) and without (blue) a leading track-$p_{\rm T}$ selection of $5~{\rm GeV}/c$, in the $10 < p_{\rm T}^{\rm ch~jet} < 15~{\rm GeV}/c$ interval. Comparisons to PYTHIA 8 simulations and pQCD predictions are also shown \cite{eec}.} 
\label{fig:eec}
\end{figure}

The EEC measurements are shown in Fig.~\ref{fig:eec}. A comparison between the inclusive-jet EEC, constructed from predominantly ($\sim76\%$) gluon-initiated jets \cite{eec}, and the light-quark jet prediction from PYTHIA 8, where the dead-cone is negligible, shows a peak shift to small $R_{\rm L}$ corresponding to narrowing of the jet core due to an effect of Casimir factors. In the $\rm D^0$-tagged EEC, one sees a reduced total integral and a significant suppression across the whole range in $R_{\textrm{L}}$. This is due to the large dead-cone in charm jets, which leads to a jet that contains fewer particles. Additionally, the peak positions of the $\rm D^0$-tagged and inclusive EECs are similar. This is due to competing effects. On one hand, the larger Casimir color factor of gluons leads to the broadening of inclusive jets that shift the inclusive EEC to larger $R_{\rm L}$. On the other hand, the dead-cone suppresses small-angle splittings and shifts the $\rm D^0$-tagged EEC toward large $R_{\rm L}$. The perturbative QCD (pQCD) calculations reproduce the general shape of the EEC but there is some tension. However, predictions from PYTHIA 8 describe the inclusive and $\rm D^0$-tagged measurements within uncertainties.

\section{$\textrm{D}^0$-tagged jet axes differences at $\sqrt{s}=5.02$ TeV}\label{sec:jetaxis}
Jet axes differences are defined as the opening angle between jet axes \cite{jetaxes}, 
\begin{equation}
\Delta R_{\rm axis} = \sqrt{\Delta \varphi^2 + \Delta \eta^2}.
\end{equation}
Several jet-axis definitions with varying sensitivity to soft radiation are studied. First, the Standard (STD) axis is obtained from the $E$-recombination scheme, which calculates the jet axis from the sum of the constituents' four-momenta. This axis is the most sensitive to soft radiation. The second is the Soft Drop (SD) groomed axis, which is created by reclustering anti-$k_{\rm T}$ jets with the Cambridge--Aachen (CA) algorithm and ordering the shower by angle. We then define the jet axis from the first splitting that passes the Soft Drop \cite{SD} condition,
\begin{equation}
\frac{min(p_{\rm T_1}, p_{\rm T_2})}{p_{\rm T_1}+p_{\rm T_2}} > z_{\rm cut} \left( \frac{\Delta R_{12}}{R} \right) ^\beta,
\label{eq:SD}
\end{equation}
with $\beta=0$ such that radiation softer than the value of $z_{\rm cut}$, a tuneable parameter, is removed. Eq.~\ref{eq:SD} is checked at each splitting in the shower, following the harder prong. Lastly, the Winner-Takes-All (WTA) jet axis is calculated from the anti-$k_{\textrm{T}}$ jets reclustered with the CA algorithm, but now utilizing the Winner-Takes-All recombination scheme so that the jet axis is defined by following the hardest subjet at each clustering step \cite{wta}. This jet axis is completely insensitive to soft radiation.

\begin{figure}
\centerline{\includegraphics[width=0.5\linewidth]{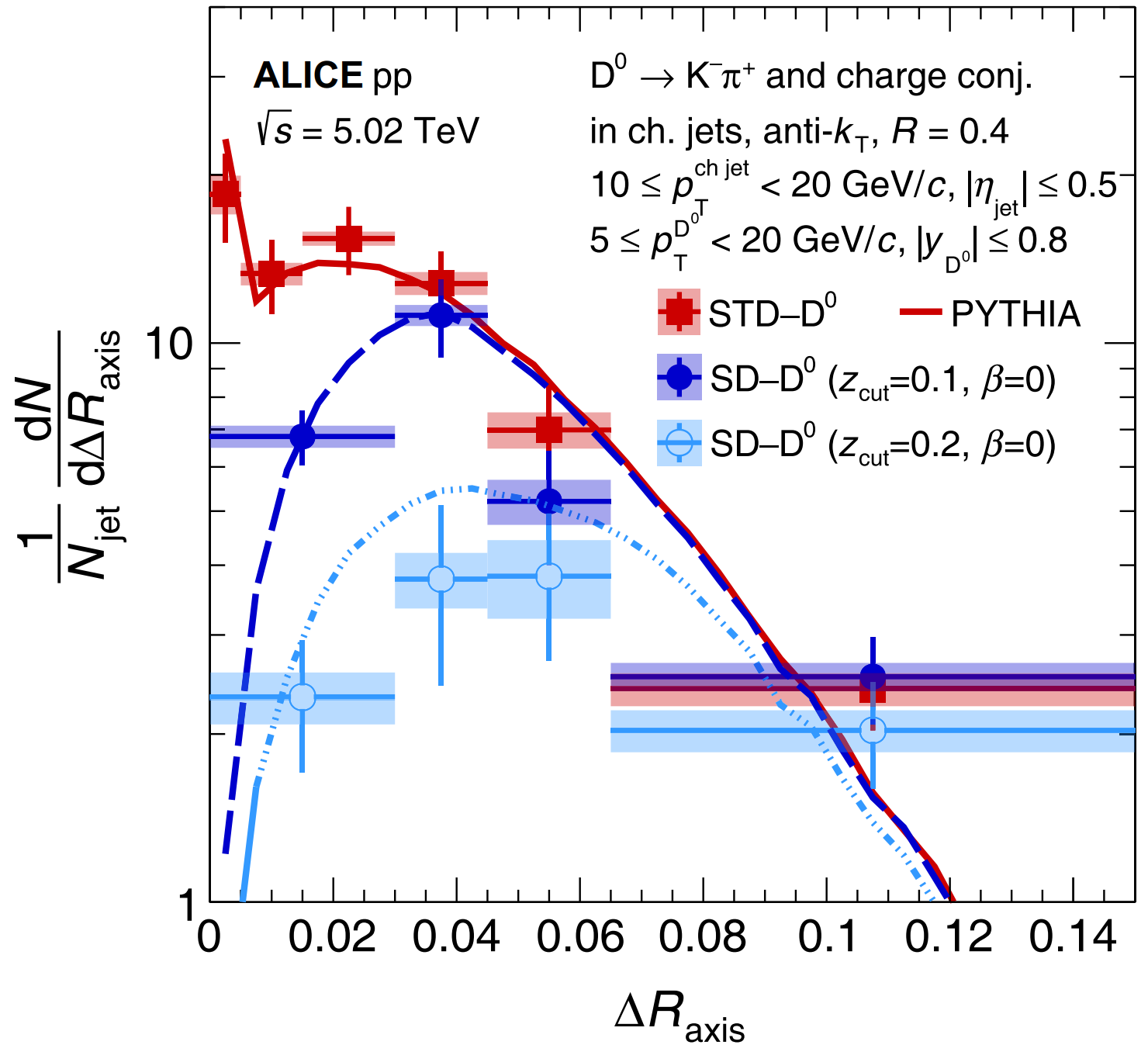}}
\caption[]{$\rm D^0$-tagged $\Delta R_{\rm STD-D^0}$ (red) compared to $\Delta R_{\rm SD-D^0}$ with $z_{\rm cut}=0.1$ (dark blue) and $z_{\rm cut}=0.2$ (light blue) in the $10 < p_{\rm T}^{\rm ch~jet} < 20~{\rm GeV}/c$ interval with a $5 < p_{\rm T}^{\rm D^0} < 20~{\rm GeV}/c$ selection applied. The measurements are compared to predictions from PYTHIA 8 \cite{jetaxes}.} 
\label{fig:jetaxes}
\end{figure}

Three jet axes differences are chosen for discussion, specifically those that study the opening angle between an axis and the D$^0$-meson direction. First, in studies of the opening angle between the Winner-Takes-All axis and the D$^0$ direction, $\Delta R_{\rm WTA-D^0}$, it is found that these two axes were aligned in $(99\pm 1)\%$ of the jets in $10 < p_{\rm T}^{\rm ch~jet} < 20~{\rm GeV}/c$ and $p_{\rm T}^{\rm D^0} > 5~{\rm GeV}/c$. Because the Winner-Takes-All axis is strongly correlated with the leading particle in the jet, this alignment implies that the $\rm D^0$ meson is the leading particle, or in the hardest prong, in all $\rm D^0$-tagged jets in the studied kinematic range. 

In Fig.~\ref{fig:jetaxes}, $\Delta R_{\rm STD-D^0}$ is compared to $\Delta R_{\rm SD-D^0}$ ($z_{\rm cut} = 0.1,0.2$). The PYTHIA prediction for $\Delta R_{\rm STD-D^0}$ shows a double-peak structure, with a broad peak near $\Delta R=0.02$. This broad peak contains jets with a charm quark that has radiated, but only very softly, such that the direction of the Standard axis is barely affected and remains close to the charm quark, the dominant track. The sharper peak observed in the first interval of $\Delta R_{\rm STD-D^0}$ comes from single-track jets, where either the charm hadron does not radiate at all, or the radiation is not captured within the jet cone. 

As a result, when the Soft Drop condition is applied, the jets that occupy small $\Delta R_{\rm STD-D^0}$ are more likely to be groomed away. Fewer jets satisfy Eq.~\ref{eq:SD} at small $\Delta R_{\rm SD-D^0}$ when $z_{\rm cut}=0.1$, and fewer still when the grooming condition is stricter ($z_{\rm cut}=0.2$). However, at the largest interval of $\Delta R_{\rm axis}$, all distributions match within uncertainties. Grooming has a minimal effect here because, in order for a jet to occupy this region, a harder, wide-angle splitting must exist that pulls the jet axis away from the core of the jet and away from the D$^0$ meson. We also see that PYTHIA reproduces the results within uncertainties, including when the grooming intensity is increased.

\section{$\textrm{D}^0$-tagged jet angularities at $\sqrt{s}=5.02$ TeV}\label{sec:ang}
The generalized jet angularities are defined as the momentum fraction of each jet constituent weighted by its angle of emission,
\begin{equation}
\lambda_{\alpha}^{\kappa} \equiv \sum_{i \in \textrm{jet}} \left ( \frac{p_{\textrm{T},i}}{p_{\textrm{T}}^{\textrm{ch jet}}} \right ) ^\kappa \left ( \frac{\Delta R_{i, \textrm{axis}}}{R} \right ) ^\alpha \equiv \sum_{i \in \textrm{jet}} z_i^\kappa \theta_i^\alpha.
\end{equation}
Here, $p_{\textrm{T},i}$ is the transverse momentum of each individual track in the jet cone, and $\Delta R_{i, \textrm{axis}}$ is the opening angle between that particle and the $E$-scheme jet axis. The parameters $\kappa$ and $\alpha$ are tuneable, but the observable is IRC safe when $\kappa=1$ and $\alpha>0$. The sum is taken over all particles in the jet, while the varied parameter, $\alpha$, allows one to highlight either collinear or wide-angle radiation. The results for $\kappa=1$ and $\alpha=1$, which emphasize collinear radiation, and the results for $\kappa=1$ and $\alpha=3$, which emphasizes wide-angle radiation \cite{ang}, are discussed.

\begin{figure}
\centerline{\includegraphics[width=0.4\linewidth]{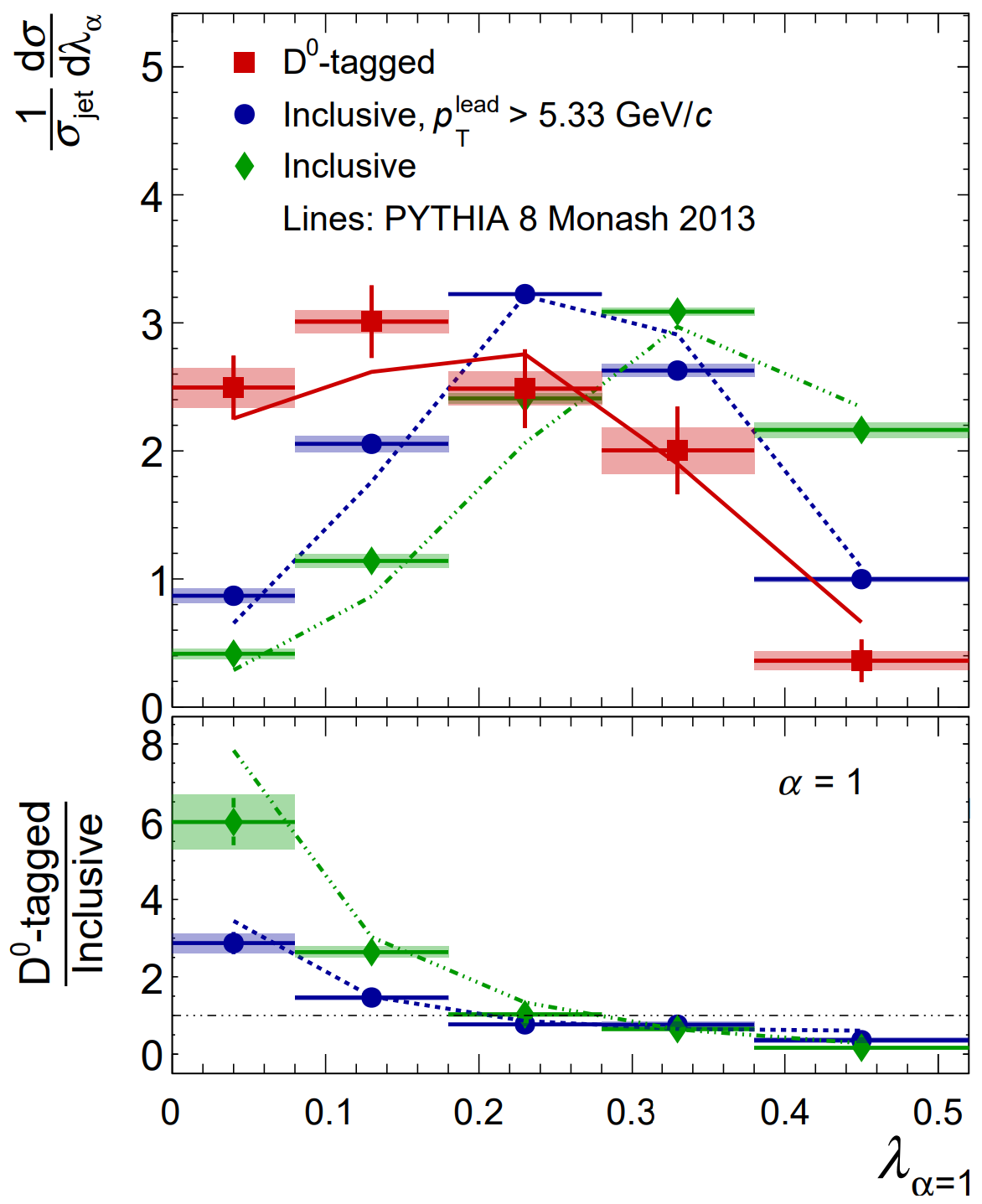}\includegraphics[width=0.4\linewidth]{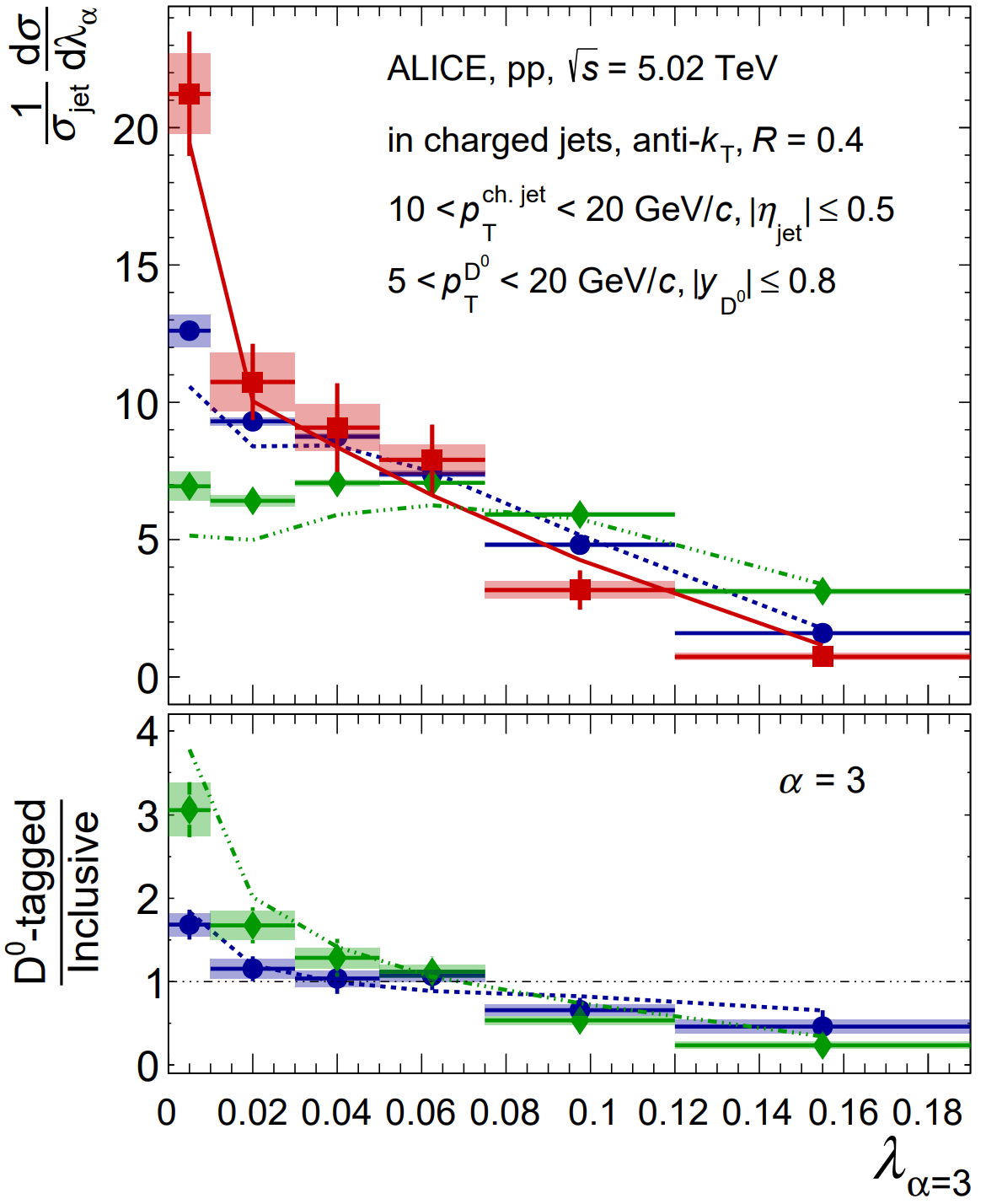}}
\caption[]{The $\rm D^0$-tagged jet angularity (red) compared to inclusive jets, with (blue) and without (green) a leading track-$p_{\rm T}$ selection of $5.33~{\rm GeV}/c$, in the $10 < p_{\rm T}^{\rm ch~jet} < 20~{\rm GeV}/c$ interval. The $\lambda_{\alpha=1}$ angularity is on left; the $\lambda_{\alpha=3}$ angularity is on right. Comparisons to PYTHIA 8 simulations are also shown \cite{ang}.} 
\label{fig:ang}
\end{figure}

The first jet angularity, $\lambda_{\alpha=1}$, is shown in Fig.~\ref{fig:ang}(a). The D$^0$-tagged angularity peaks at significantly lower values than the gluon-dominated, inclusive angularity. This implies that the transverse momentum carried by hadrons is concentrated closer to the jet axis in D$^0$-tagged jets, whereas it is distributed to wider angles in inclusive jets. This difference can be attributed to the fact that jets tagged with a heavy quark generally have a more concentrated jet core than gluon-dominated jets. It may immediately be counterintuitive that the charm-tagged distribution peaks at small angularities given a depopulated collinear region (dead-cone). However, when one considers the angular term is weighted by the momentum fraction of each particle, it is evident that very soft radiation would likely sit at smaller angularity, which overpowers the angular widening from the dead-cone. These results are also compared with an inclusive angularity that includes a selection on the leading-track $p_{\textrm{T}}$. This selection enhances the number of quark jets in the sample, specifically light-quark-initiated jets. As seen in the ratios in Fig.~\ref{fig:ang}, this angularity is similar to the D$^0$-tagged result but remains distinct from the inclusive angularity, due to the onset of color factors. As shown in Fig.~\ref{fig:ang}(b), while trends between D$^0$-tagged and inclusive angularities persist into $\lambda_{\alpha=3}$, the differences are reduced compared to $\lambda_{\alpha=1}$ due to the increased emphasis given to wide-angle radiation, where the dead-cone effect is less prominent and sensitivities to color-charge differences are enhanced.

\section{Conclusion}
Heavy-flavor jets are an ideal probe for studying jet evolution and fragmentation, revealing the interplay of the QCD dead cone and Casimir color factors in charm-initiated parton showers. The low energies studied at ALICE give access to a prominent dead-cone, which causes a reduction in the number of parton splittings, lowering the EEC amplitude and increasing the contribution of single-track jets. Despite the large dead-cone, jets initiated by heavy-quarks remain narrower due to the effects of Casimir color factors. We have validated that the radiation pattern of charm quarks is subject to these QCD phenomena in pp collisions, and that modeling the parton shower with PYTHIA reproduces the observed jet features in this kinematic regime.

\end{document}